\documentclass[aps,10pt,twocolumn,showpacs,preprintnumbers,amsmath,amssymb,prb,floatfix]{revtex4-1}
\usepackage{amsmath,amssymb}
\usepackage{graphicx}
\usepackage{verbatim}

\usepackage{color}

\begin{document}

\title{Optical bistability of graphene in the terahertz range}

\author{N. M. R. Peres$^{1}$}
\author{Yu. V. Bludov$^{1}$}
\author{Jaime E. Santos$^{1}$}
\author{Antti-Pekka Jauho$^{2}$}
\author{M. I. Vasilevskiy$^{1}$}

\affiliation{$^{1}$Centro de F\'{\i}sica and Departamento de F\'{\i}sica, Universidade
do Minho, Campus de Gualtar, Braga 4710-057, Portugal}
\affiliation{$^{2}$Center for Nanostructured Graphene (CNG), Department of Micro and Nanotechnology,
Technical University of Denmark, DK-2800 Kongens Lyngby, Denmark}

\begin{abstract}
We use an exact solution of the relaxation-time Boltzmann equation in an uniform AC electric field
to describe the nonlinear optical response of graphene in the terahertz (THz). 
The cases of monolayer, bilayer and ABA-stacked trilayer graphene are considered, 
and the monolayer species is shown to be the most appropriate one to exploit the 
nonlinear free electron response. We find that a single
layer of graphene shows  optical bistability in the THz range, 
within the electromagnetic power range attainable in practice.
The current associated with the third harmonic generation is also computed.
\end{abstract}

\pacs{42.65.Wi, 78.67.Wj, 73.25.+i, 78.68.+m}

\maketitle

\section{Introduction}
Optical bistability is a way of controlling light with light \cite{Gibbs,Shen82}.
Bistability refers to an optical effect where a system exhibits two
different values of the transmitted light intensity for a single value of the input intensity.
One way of analyzing the bistability is to explore the optical Kerr-effect, 
a non-linear phenomenon where the light the modulates material's refractive index
\cite{Boyd}.
In general, for the effect to be measurable, the light field must transverse a macroscopic
distance within the non-linear material. In a semiconductor, optical bistability was observed
a long time ago \cite{Gibbs79}. The desired goal in the field of optical bistability 
is the possibility of realizing in a single device a set of functionalities,
such as switching, logic functions, memory with a fast time response, and modulation,
all using a low power laser \cite{Almeida04}. Eventually, the practical realisation
of an optical computer is in the horizon \cite{Smith85}.

In general, optical bistability can be realized at the interface between a linear and
a non-linear material, with the reflected light intensity showing hysteresis
\cite{Smith79}. However, what may seem surprising is that the hysteresis  can be observed
in a system one-atom thick, such as graphene. In the optical region of the spectrum,
it has been shown that graphene has a strong non-linear optical response
\cite{Hendry10,Gu12,Kim12,Mikhailov12}.
The same phenomenon has been observed in graphene derivatives \cite{Liaros13} and in graphene
nano-ribbons intercalated with boron nitride \cite{Zhang14}. It has also been shown
that graphene can dramatically change the nonlinear response of
a silicon photonic crystal \cite{Kim12}.

Theoretically, the non-linear response of graphene at optical frequencies has been
exploited to produce a novel class of nonlinear self-confined modes
\cite{Nesterov}. On the other hand,
in the THz spectral range, graphene has the potential for many
applications \cite{Bludov2013,Low14,Abajo14,Stauber14}.

Some aspects of the non-linear optical properties of graphene have already been considered
in the literature \cite{Mikhailov_2007,Mikhailov_2008,Mikhailov_2008_b,Mikhailov_2009}.
However, the exploitation of those properties to the problem of bistability was not considered
before. Results for the non-linear Drude response of graphene
in the collisionless regime have been derived previously
\cite{Mikhailov_2007,Mikhailov_2008,Mikhailov_2008_b,Mikhailov_2009}.
Here we extend the derivation to the regime where a finite relaxation time exists,
given two alternative methods to generate the expansion (one of them non-perturbative).
The response of graphene to an
electromagnetic pulse has also been obtained \cite{Mikhailov_2008_b}.
It has also been shown that strong magnetic fields, which drive the system to the quantum Hall
regime, can induce a {\it giant optical non-linearity} in graphene \cite{Yao_2012}.
In addition, the latter authors, have also discussed
an efficient nonlinear generation of THz plasmons in graphene \cite{Yao_2014}.

In this paper
we show that graphene has a strong non-linear response in the THz leading to the
phenomenon of bistability. This property may allow the fabrication of active devices in this
spectral range. Furthermore, the study of non-linear surface plasmon polaritons on graphene
becomes accessible, since we can now solve the dispersion relation in the presence of a
field-dependent conductivity. Indeed, one can even envision controlling light with light
exploiting plasmonic nanostructures \cite{Martti12}.

The article is organized as follows. In Sec. \ref{sec:Boltzmann} 
we present the general solution of the Boltzmann equation, which is exact
 within the momentum-independent relaxation time approximation.
This solution is used in Sec. \ref{sec:Conductivity} to calculate the
 frequency-dependent nonlinear conductivity of monolayer, bilayer, 
and ABA-stacked trilayer graphene. The THz optical bistability in
 monolayer graphene is considered in Sec. \ref{sec:Bistability} and the 
last section is devoted to conclusions.

\section{Boltzmann equation for a 2D electron system under AC electric field}
\label{sec:Boltzmann}

{In the presence of an AC field, $\mathbf E= E(t)\mathbf u_x$ (which is directed along $x$-axis and the time dependence of $E(t)$, in principle, can have an arbitrary form), within the relaxation time approximation, the Boltzmann equation reads:}
\begin{equation}
 \frac{\partial f_n({\mathbf k},t)}{\partial t} -\frac{e}{\hbar} E(t)\frac{\partial f_n({\mathbf k},t)}{\partial k_x}
=-\frac{f_n({\mathbf k},t)-f_0[\epsilon_n(\mathbf k)]}{\tau}\,,
\label{BE}
\end{equation}
where $f_0[\epsilon_n(\mathbf k)]$ is the Fermi-Dirac distribution function, {$\epsilon_n(\mathbf k)$ is the $n$-th band energy of 2D electrons with ${\mathbf k}=(k_x,k_y)$,} and $\tau$ is the (microscopic) relaxation time. As shown in Appendix \ref{sec:AppA}, this equation can be solved analytically if we assume that the microscopic relaxation time does not depend on $\mathbf k$. Although it might look unrealistic at first sight, this approximation is justified by the fact that $\tau $ disappears from the expression for the electric current in the physically interesting limit of frequencies ($\omega \tau \gg 1$), as it will be shown below. Alternatively, 
one can solve Eq. (\ref{BE}) by iterations (see Appendix \ref{sec:AppB}), 
a procedure that allows to take into account the dependence of the 
microscopic relaxation time upon the electron momentum.
The exact solution is:
\begin{equation}
f_n(\mathbf k,t) = e^{-t/\tau}\int_{-\infty}^t\frac{dt'}{\tau}e^{t'/\tau}
f_0[\epsilon(k_x+\kappa(t,t'),k_y)]\,.
\label{BeqSol}
\end{equation}
Here we introduced a shorthand notation $\kappa(t,t')=(e/\hbar)\int^{t}_{t'}E(t'')dt''$.
For a harmonic time-dependence $E(t) = E_0\cos(\omega t)$, which will be focus of our study, the function $\kappa$ is
\begin{equation}
 \kappa(t,t')=\int^t_{t'}\frac{eE(t'')}{\hbar}dt''=\frac{eE_0}{\hbar \omega}[\sin(\omega t)-\sin(\omega t')]\,.
\end{equation}
\section{Non-linear current response}
\label{sec:Conductivity}

\subsection{General expression}

The current is given in terms of the solution of the Boltzmann equation,
Eq. \eqref{BeqSol}, by
\begin{eqnarray}
 j_x&=&-\frac{e}{\pi^2\hbar}\sum_{n=1}^N\int d\mathbf k\frac{\partial \epsilon_n}{\partial k_x}\, f_n(\mathbf k,t)\nonumber\\
    &=& -\frac{e}{\pi^2\hbar}
e^{-t/\tau}\int^t_{-\infty}\frac{dt'}{\tau}e^{t'/\tau}\nonumber\\
&\quad&\times\sum_{n=1}^N\int d\mathbf k\frac{\partial \epsilon_n}{\partial k_x} f_0[\epsilon_n(k_x+\kappa(t,t'),k_y)]\,,\label{eq:jx}
\end{eqnarray}
where $N$ is the number of bands in the spectrum (e.g. 2 in the case of bilayer graphene) and the integration is over the first Brillouin zone.
In the low temperature limit ($T\rightarrow 0$) the equilibrium Fermi-Dirac distribution function can be replaced by the Heaviside step-function $\theta$, so that the non-equilibrium distribution function becomes:
\begin{equation}
f_0[\epsilon_n(k_x+\kappa(t,t'),k_y)]=\theta[\epsilon_F-\epsilon_n(k_x+\kappa(t,t'),k_y)].\label{eq:fnoneq}
\end{equation}
In the following, 
Eqs.(\ref{eq:jx}--\ref{eq:fnoneq}) will be used to
 compute the non-linear response in different forms of
 graphene where the electronic energy spectra are different.

\subsection{Monolayer graphene}

The spectrum of monolayer graphene consists of only one band ($N=1$), which in the Dirac cone approximation can be respresented as $\epsilon_1(\mathbf k)=v_F\hbar\sqrt{k_x^2+k_y^2}$  ($v_F=\sqrt{3}at/(2\hbar)$ is the Fermi velocity of the electrons,  $a$  is the lattice constant and $t$ is the tight-binding nearest-neghbour hopping parameter). To compute $j_x$ we first focus our attention on the momentum integration.
To that end, we define the integral
\begin{equation}
I_{11}(\kappa)=\int dk_xdk_y\frac{k_x}{\sqrt{k_x^2+k_y^2}}\theta \left (\epsilon_F-v_F\hbar\sqrt{(k_x+\kappa)^2+k_y^2}\right )\,,
\label{eq:I11}
\end{equation}
such that
\begin{equation}
j_x=-\frac{ev_F}{\pi^2}e^{-t/\tau}\int^t_{-\infty}\frac{dt'}{\tau}e^{t'/\tau}I_{11}(\kappa),
\end{equation}
and $\kappa\equiv \kappa(t,t')$.  Note that we consider a doped graphene sheet, i.e., we assume a finite $\epsilon_F$ (and
a corresponding finite $k_F=\epsilon_F/(\hbar v_F)$).
Performing the substitutions $k_x+\kappa=\tilde k_x$, $k_y=\tilde k_y$
the integral becomes
\begin{equation}
 I_{11}(\kappa)=\int d\tilde k_xd\tilde k_y\frac{\tilde k_x-\kappa}{\sqrt{(\tilde k_x-\kappa)^2+\tilde k_y^2}}
\theta(\epsilon_F-v_F\hbar\sqrt{\tilde k_x^2+\tilde k_y^2})\,.\label{eq:I11-sim}
\end{equation}
\begin{figure}[!th]
\includegraphics*[width=8cm]{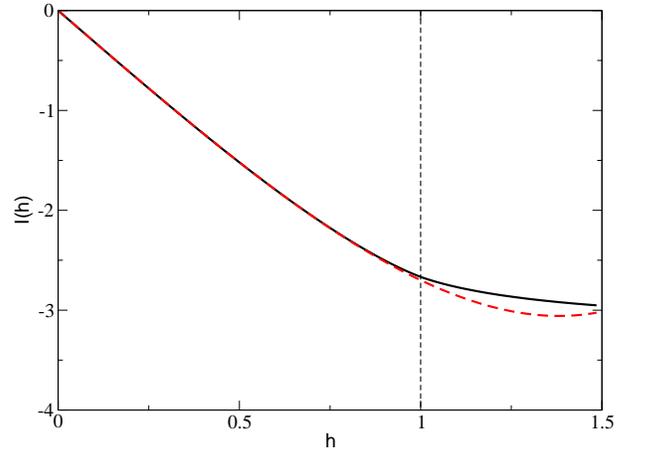}
\caption{(color online) Plot of the function $I(\kappa)$ {\it vs} $h=\kappa / k_F$.
Function $I(\kappa)$, as computed from Eq. (\ref{eq_elliptic}), and (dashed line)
the approximation  given by Eq. (\ref{eq_expansion}); we have taken $k_F=1$.}
\label{fig_expansion}
\end{figure}
Introducing the limits of integration imposed by the step-function,
the integral splits into two terms
\begin{eqnarray}
& &I_{11}(\kappa)=\int_{-k_F}^{k_F}d\tilde k_y\int_0^{\sqrt{k_F^2-k_y^2}}
d\tilde k_x \frac{\tilde k_x-\kappa}{\sqrt{(\tilde k_x-\kappa)^2+\tilde k_y^2}}+\nonumber\\
& &\int_{-k_F}^{k_F}d\tilde k_y\int^0_{-\sqrt{k_F^2-k_y^2}}
d\tilde k_x \frac{\tilde k_x-\kappa}{\sqrt{(\tilde k_x-\kappa)^2+\tilde k_y^2}}\,.
\end{eqnarray}
The integral over $d\tilde k_x$ is elementary and we end up with
\begin{eqnarray}
 I_{11}(\kappa)&=&2\int_0^{k_F}d\tilde k_y
\left(\sqrt{\kappa^2+k_F^2-2\kappa\sqrt{k_F^2-\tilde k_y^2}}\right.\nonumber\\
&\quad&-\left.\sqrt{\kappa^2+k_F^2+2\kappa\sqrt{k_F^2-\tilde k_y^2}}\right)\,.
\label{eq_elliptic}
\end{eqnarray}
We note that $I_{11}(\kappa)$ is an odd function of $\kappa$. The integral $I_{11}(\kappa)$
can be written in terms of elliptic integrals, a result valid for all
values of the ratio $\kappa/k_F$.
In the regime $\kappa/k_F\le1$ the integral $I(\kappa)$
can be expressed in terms of the Gaussian hypergeometric function\cite{Abramowitz}
$_2F_1(a,b;c,x)$ as
\begin{equation}
 I_{11}(\kappa)=-\pi \kappa k_F\times\,_2F_1\left(-\frac{1}{2},\frac{1}{2};2,\frac{\kappa^2}{k_F^2}\right)\,.
\label{eq_elliptic_exact}
\end{equation}
Although this is a formal analytical expression, it is
preferable to expand it in powers of $\kappa/k_F$,
\begin{eqnarray}
 & & I_{11}(\kappa)\approx-\pi k_F \kappa \left [1 - \frac{1}{8}\left (\frac{\kappa}{k_F} \right )^2 - \frac{1}{64}\left (\frac{\kappa}{k_F} \right )^4 \right ]\,.
\label{eq_expansion}
\end{eqnarray}
It is important to stress that all terms but the first in this series have the same sign. A comparison between the
result of Eq. (\ref{eq_elliptic}) with the approximate expression (\ref{eq_expansion})
is given in  Fig. \ref{fig_expansion}. Clearly, the expansion
(\ref{eq_expansion}) works very well all the way from $\kappa/k_F=0$ till $\kappa/k_F=1$.

To evaluate the current at zero temperature
we still need to compute the integral over $t'$. The first-order term is
\begin{eqnarray}
J^{(1)}&=&e^{-t/\tau}\int^t_{-\infty}\frac{dt'}{\tau}e^{t'/\tau}\kappa(t,t')
\nonumber\\
&\quad&=
\frac{eE_0}{\hbar}\frac{\tau[\cos(t\omega)+\tau\omega\sin(t\omega)]}{1+\tau^2\omega^2}\nonumber\\
&\quad&=\frac{eE_0}{2\hbar}\tau\frac{1+i\tau\omega}{1+\tau^2\omega^2}e^{-i\omega t}+{\rm c.c.}\,.\label{eq:J1}
\end{eqnarray}
The current is thus
\begin{equation}
 j_x^{(1)}=\frac{ev_Fk_F}{\pi}J^{(1)}=\frac{e^2}{\pi\hbar}\frac{\epsilon_F\tau}{\hbar}
\frac{1}{1-i\tau\omega}\frac{E_0}{2}e^{-i\omega t}\,,\label{eq:monolay_sig}
\end{equation}
which is nothing but Drude's result. Here we have extracted the dependence of the integral on $e^{-i\omega t}$ only. In the limit $\omega\tau\gg 1$, the linear part of the current can be expressed as
\begin{equation}
 j_x^{(1)}=i\nu_1\frac{E_0}{2}e^{-i\omega t}\,, \qquad
\nu_1=\frac{e^2}{\pi\hbar}\frac{\epsilon_F}{\hbar\omega}\:.
\label{eq:nu1}
\end{equation}

The calculation of the third order term is more tedious.\cite {note1} We have to evaluate
\begin{eqnarray}
&&J^{(3)}=e^{-t/\tau}\int^t_{-\infty}\frac{dt'}{\tau}e^{t'/\tau}\kappa^3(t,t')=\nonumber\\
 &&-\frac{18\tau^3}{(2i\tau\omega-1)(1+\tau^2\omega^2)}
\left(\frac{eE_0}{2\hbar}\right)^3e^{-i\omega t}\\
&&+ \frac{6\tau^3}{1-6i\omega\tau-11\tau^2\omega^2+6i\tau^3\omega^3}
\left(\frac{eE_0}{2\hbar}\right)^3e^{-i3\omega t}+{\rm c.c.}\,\nonumber
\end{eqnarray}
which for $\omega \tau \gg 1$ leads to
\begin{equation}
 J^{(3)}=\frac{9i}{\omega^3}\left(\frac{eE_0}{2\hbar}\right)^3e^{-i\omega t}-
\frac{i}{\omega^3}\left(\frac{eE_0}{2\hbar}\right)^3e^{-i3\omega t}\,.
\end{equation}
The third order current, $j_x^{(3)}$, is given by
\begin{equation}
 j_x^{(3)}=-\frac{ev_F}{8\pi k_F}J^{(3)}= j_x^{(3,\omega)}+j_x^{(3,3\omega)}\,,\label{eq:3-h}
\end{equation}
 where
\begin{equation}
 j_x^{(3,\omega)} = -i\nu_3\frac{E_0^3}{8}e^{-i\omega t}\,,
\qquad \nu_3=9\frac{e^2}{\hbar \pi}\frac{v_F^2}{8\epsilon_F}\frac{e^2}{\hbar\omega^3}\label{eq:nu3}
\end{equation}
and
\begin{equation}
  j_x^{(3,3\omega)} = i\frac{\nu_3}{9}\frac{E_0^3}{8}e^{-i3\omega t}\,.
  \label {j3-2}
\end{equation}
The term $j_x^{(3,3\omega)}$ represents the third
harmonic generation. We also note that  result (\ref{eq:nu3}) differs by a factor of 3
from the result for the same quantity computed by Mikhailov \cite{Mikhailov_2007}.
This difference exists, because Mikhailov treatment does not permit to study the 
regime of $\omega\tau\gg1$, since by construction it assumes that
the observation time is much smaller than $\tau$.
Finally, the current to fifth order (in the limit $\tau\omega\gg1$) is given
by
\begin{equation}
 j_x^{(5,\omega)}=-i\nu_5\left(\frac{E_0}{2}\right)^5e^{-i\omega t}\,,
\qquad \nu_5=\frac{25}{16}\frac{e^2}{\hbar\pi}
\frac{v_F^4}{\epsilon_F^3}\frac{e^4}{\hbar\omega^5}.\label{eq:5-h}
\end{equation}
This concludes the derivation of the nonlinear response of monolayer graphene.
An alternative way to obtain the nonlinear current in monolayer graphene is
presented in Appendix \ref{sec:AppB}, where the Boltzmann equation is solved
by means of expansion of the nonlinear distribution function in powers
 of the electric field, while here the expansion was
performed during the calculation of the current density. In both
 cases the dimensionless expansion parameter is $k_0/k_F$,
where $k_0=eE_0/(\hbar\omega)$. The procedure is valid if $k_0/k_F<1$.

\subsection{Bilayer and trilayer graphene}

We next consider a AB-stacked graphene bilayer, 
whose spectrum  consists of two parabolic bands ($N=2$) 
and can be represented as\cite{c:bilayer}
\begin{eqnarray}
& &\epsilon_1({\bf k})=\frac{v_F^2\hbar^2(k_x^2+k_y^2)}{t_\perp}\:;\label{eq:e1}\\
& &\epsilon_2({\bf k})=t_\perp+\frac{v_F^2\hbar^2(k_x^2+k_y^2)}{t_\perp},\label{eq:e2}
\end{eqnarray}
where $t_\perp$ is the hopping parameter between the layers. Substituting (\ref{eq:e1}) and (\ref{eq:e2}) into Eq. (\ref{eq:jx}), we obtain the following expression for the current density:
\begin{eqnarray}
j_x=-\frac{2ev_F^2\hbar}{\pi^2t_\perp}e^{-t/\tau}\int^t_{-\infty}\frac{dt'}{\tau}e^{t'/\tau}\left[I_{21}(\kappa)+I_{22}(\kappa)\right],
\label{eq:jx2}
\end{eqnarray}
where $I_{2n}$ with $n=1,\:2$ are  integrals analogous to $I_{11}$ defined in the previous section, and
are evaluated in Appendix \ref{sec:AppC}.
Substituting them into Eq. (\ref{eq:jx2}) and using Eq. (\ref{eq:J1}) we obtain:
\begin{eqnarray}
& &j_x=\frac{2e}{\pi\hbar }\left[\epsilon_F+\left(\epsilon_F-t_\perp\right)\theta\left(\epsilon_F-t_\perp\right)\right]J^{(1)}\nonumber\\
& &=\frac{2e^2}{\pi\hbar^2 }\left[\epsilon_F+\left(\epsilon_F-t_\perp\right)\theta\left(\epsilon_F-t_\perp\right)\right]\frac{E_0}{2}\tau\frac{1+i\tau\omega}{1+\tau^2\omega^2}e^{-i\omega t}.\nonumber\\\label{eq:bilay_sig}
\end{eqnarray}
It is interesting that, owing to its parabolic energy spectrum [Eqs. (\ref{eq:e1})--(\ref{eq:e2})], bilayer graphene is a purely linear system. If the Fermi level is below the interlayer hopping energy $t_\perp$, the conductivity is equal to twice the first order conductivity of monolayer graphene [compare Eqs. (\ref{eq:bilay_sig}) and (\ref{eq:monolay_sig})]. For $\epsilon_F>t_\perp$, there is a correction to the conductivity due to the second band filling.


The spectrum of the ABA-stacked trilayer graphene consists of one Dirac-type and two parabolic bands,\cite{c:trilayer}
\begin{eqnarray}
\label{eq:parab}
& &\epsilon_{1}=\frac{\hbar^2v_F^2(k_x^2+k_y^2)}{\sqrt{2}t_\perp}\:;\\
& &\epsilon_{2}=\hbar v_F\sqrt{k_x^2+k_y^2}\:;\\
& &\epsilon_{3}=\sqrt{2}t_\perp+\frac{\hbar^2v_F^2(k_x^2+k_y^2)}{\sqrt{2}t_\perp}\:.
\end{eqnarray}
Substituting these relations into Eq. (\ref{eq:jx}) and proceeding as before we obtain the following expression for the induced current:
\begin{eqnarray}
\nonumber
& & j_x=\frac{e^2}{\pi\hbar^2 }\left[3\epsilon_F+2\left(\epsilon_F-\sqrt{2}t_\perp\right)\theta\left(\epsilon_F-\sqrt{2}t_\perp\right)\right]\times\\
& & \frac{E_0}{2}\tau\frac{1+i\tau\omega}{1+\tau^2\omega^2}e^{-i\omega t}+j_x^{(3)}+j_x^{(5,\omega)}\:.
\end{eqnarray}
Here $j_x^{(3)}$, $j_x^{(5,\omega)}$ coincide with those defined by Eqs. (\ref{eq:3-h}) and (\ref{eq:5-h}), respectively. The main result is that, in contrast with the case of bilayer graphene, this material is a nonlinear medium alike monolayer graphene. However, the linear part of the induced current in this case is larger than for monolayer graphene, so we may say that its nonlinearity is relatively weaker.

We note that below we use the expressions for the non-linear optical response of graphene
in the collisionless regime. This may be experimentally justified. In a previous experimental
study \cite{Lee_2011}
of the transmittance of graphene in the  wavenumber range of $[30, 1000]$ cm$^{-1}$,
a relaxation rate of $\Gamma=95$ cm$^{-1}$ was found (see Fig. 3 of that reference).
For the two frequencies considered below the product $\omega\tau=2\pi f\tau$ is $2.2$ and $1.1$
for the frequencies of $f=1$ THz and $f=0.5$ THz, respectively
(see also Ref. \onlinecite{Mak_2012} for different (smaller) values of $\Gamma$). Clearly these numbers are not in
the in the regime $\omega\tau\gg1$. However, these numbers are for large area CVD grown graphene,
which is known to produce a low-mobility material. On the other hand, exfoliated graphene has
mobilities that are more than one order of magnitude larger.
In an experiment done in this type of graphene
one would be in the regime $\omega\tau\gg1$.
Indeed, a recent  theoretical calculation\cite{Sule_2014} of the optical response of suspended graphene in the
terahertz range,
using ab-initio methods,
 yielded a value of $\Gamma=1/\tau\sim0.8$ THz which leads to $2\pi f/\Gamma\sim 7.9$.
\section{Bistability of monolayer graphene}
\label{sec:Bistability}

\begin{figure}[!th]
\includegraphics*[width=5cm]{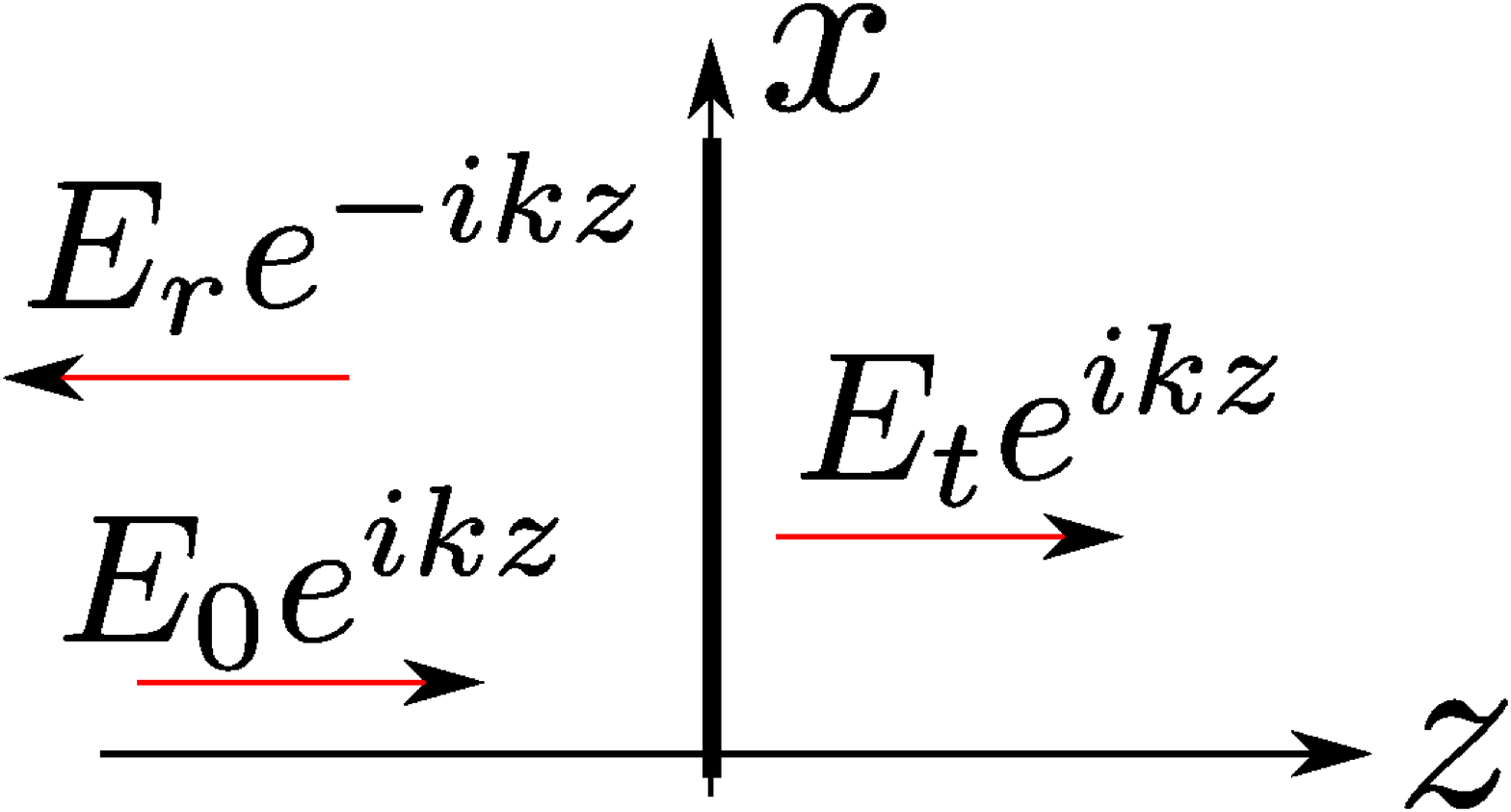}
\caption{(color online) Scattering geometry. The thick line represents the graphene sheet.}
\label{fig_scatter}
\end{figure}

We shall now discuss the possibility of optical bistability in graphene. To this end,
we start by solving the scattering problem in the geometry defined by
Fig. \ref{fig_scatter}, where a graphene sheet, the non-linear medium, is located at
$z=0$.
The boundary conditions obeyed by the electromagnetic field are
\begin{equation}
 E_r+E_0=E_t\,,
\end{equation}
 and
\begin{equation}
 B_L-B_R=\mu_0j_x\,,
\end{equation}
where $B_L$ is the magnetic field of the electromagnetic field to the left of graphene and $B_R$
that to the right. From Maxwell's equations it follows that
\begin{equation}
 \partial_zE_x=i\omega B_y\,,
\end{equation}
which imply that
\begin{equation}
 B_L=\frac{k}{\omega}(E_0-E_r)\,,
\end{equation}
and
\begin{equation}
 B_R=\frac{k}{\omega}E_t\,.
\end{equation}
Thus
\begin{equation}
 \frac{k}{\omega}(E_0-E_r)-\frac{k}{\omega}E_t=i\mu_0(\nu_1E_t-\nu_3E_t^3
-\nu_5E_t^5)\,,
\end{equation}
or
\begin{equation}
 E_0=E_t\left[
1-i\frac{\mu_0c}{2}(\nu_1-\nu_3E_t^2
-\nu_5E_t^4)
\right]\,,
\label{eq_E0}
\end{equation}
where $\nu_1$, $\nu_3$, $\nu_5$ are defined in Eqs. (\ref{eq:nu1}), (\ref{eq:nu3}), and (\ref{eq:5-h}).
We must stress the bistability effect does not require the inclusion of the fifth order term.
We only include it here to show that the effect is not suppressed by higher order powers of the
expansion.
Here we suppose for convenience that $E_t$ is purely real, i.e. possesses zero phase, then $E_0$ is complex.
Taking the square of the modulus of Eq. (\ref{eq_E0}), we obtain
\begin{equation}
\vert E_0 \vert ^2=E_t^2\left[
1+ \frac{\mu_0^2c^2}{4}\nu_1^2
\left(
1-\frac{\nu_3+E_t^2\nu_5}{\nu_1}E_t^2
\right)^2
\right]\,.
\label{eq_E02}
\end{equation}
Defining $\vert E_0\vert ^2=Y$ and $E_t^2=X$ [\onlinecite{Markos}], we rewrite Eq. (\ref{eq_E02}) as
\begin{figure}[t]
\includegraphics*[width=8cm]{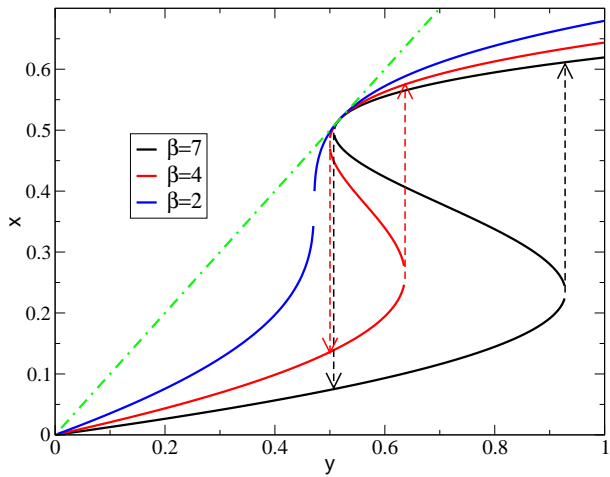}
\caption{(color online) Bistability curves of
the dimensionless field
$x$ as function of $y$, for different values of the parameter $\beta$.
When the power of the laser
is increased the transmission through graphene follows the curve starting at zero until
it reaches a point where the transmission suffers a sudden jump to higher values.
The dashed-dotted straight line is the function $x=y$.}
\label{fig_bi}
\end{figure}
\begin{equation}
Y=X \left[1+ \beta
\left(
1-\Lambda X
\right)^2
\right]
 \label{eq_X_Y}
\end{equation}
where
\begin{equation}
 \beta= \frac{\mu_0^2c^2}{4}\nu_1^2=4\alpha^2\frac{\epsilon_F^2}{\hbar^2\omega^2}\,,
\end{equation}
is a dimensionless parameter, $\alpha$ is the fine structure constant,
and
\begin{equation}
 \Lambda= \frac{\nu_3+E_t^2\nu_5}
{\nu_1}=\frac{9}{8}\frac{v_F^2\hbar^2}{\epsilon_F^2}
\frac{e^2}{\hbar^2\omega^2}+
\frac{25}{16}\frac{v_F^4}{\epsilon_F^4}\frac{e^4}{\omega^4}X\,.
\label{eq:Lambda}
\end{equation}
Clearly, it follows from Eq. (\ref{eq_X_Y}) that for $X=1/\Lambda$ resonant transmission occurs,
that is, the system becomes fully transparent ($X=Y$).\cite {note2}

It is more convenient to rewrite Eq. (\ref{eq_X_Y}) in dimensionless form. To that end
we introduce the new variables
\begin{equation}
 x=\frac{e^2E_t^2}{\hbar^2\omega^2k_F^2}
\end{equation}
and
\begin{equation}
 y=\frac{e^2E_0^2}{\hbar^2\omega^2k_F^2}\,,
\end{equation}
which leads to a universal equation for the relation between $x$ and $y$ as function of the
dimensionless parameter $\beta$:
\begin{equation}
 y=x\left[
1+\beta\left(
1-\frac{9}{8}x-\frac{25}{16}x^2
\right)^2
\right]
\label{eq_x_y}
\end{equation}

Let us now analyse the consequences of Eq. (\ref{eq_x_y}). For a given value of
$E_0$ this equation has one or more real solutions, such that $E_t<E_0$. These
solutions are depicted in Fig. \ref{fig_bi}.
From this figure we see that there is a region of incoming intensities ({$Y_-\le Y \le Y_+$})
for which there are three possible values of the transmitted intensity ($X$). However, the intermediate one corresponds to an unstable state (like in the case of first-order phase transitions).
If one starts at small values of $Y$ and cranks up the intensity of the laser,
one follows the lower curve till a point {$Y_+$} where there is a sudden jump in the
transmitted intensity $X$, represented by an arrow pointing up. On the other hand, if one starts
at a high power and reduces it,  the transmitted power will follow the solid curve,
until it suddenly jumps to a regime of low transmission {($Y_-$)}, represented by
a dashed line with an arrow pointing down. This implies that there is a
hysteresis effect, or bistability. We should emphasize that this bistability is of
electronic origin and, therefore, the switching of the bistability should be quite
fast.

The incident power domain where Eq. (\ref{eq_X_Y}) has three roots can be found by putting its discriminant equal to zero, namely
\begin{eqnarray}
27\beta\Lambda^2Y^2
-4\beta\Lambda(\beta+9)Y+4(1+\beta)^2=0\:
\label{eq:discr}
\end{eqnarray}
(the last term in (\ref{eq:Lambda}) was neglected for simplicity). From Eq. (\ref{eq:discr}) we obtain
\begin{eqnarray}
Y_\pm=\frac{2}{27\beta\Lambda}\left[\beta(\beta+9)\pm\sqrt{\beta(\beta-3)^3}\right].
\label{eq:Ypm}
\end{eqnarray}
It follows from Eq. (\ref{eq:Ypm}) that if $\beta\le 3$ there is only one root of Eq. (\ref{eq_X_Y}), i.e. there is no bistability. For $\beta>3$ an increase of $\beta$ leads to the broadening of the bistability domain $Y_-\le Y \le Y_+$.

The solution of the bistability equation in terms of dimensionless variables
allow us to control the validity of the expansion, since for the considered parameters
we always have $x<1$, that is, the condition  $h=k_0/k_F<1$ [with $k_0=eE_t/(\hbar\omega)$]
is not violated along the hysteresis curve.

\section{Conclusions}
In summary, we analysed the nonlinear response of doped monolayer and multilayer graphene in the THz range, where it is determined by intraband transitions of free electrons.
Our analysis, based on an exact solution of the relaxation-time Boltzmann equation,
shows the crucial role of the Dirac-type electronic spectrum in getting considerable (third-order)
nonlinearity
and indicates monolayer graphene as the most appropriate one to exploit it.
The nonlinearity causes the third harmonic generation (the current $ j^{(3,3\omega)} $
calculated in Sec. \ref {sec:Conductivity}) and the optical bistability considered in the
 previous section. The latter is important because of its potential for applications in THz
 laser pulse modulation, optical switching, and signal processing. The estimated switching
powers are attainable with existing terahertz radiation sources.
In fact, THz lasers with peak electric fields of $\sim 4$ MV/m have recently been
built \cite{Beck10}. Single-cycle THz pulses with amplitudes exceeding 100 MV/m
are also possible \cite{Hirori11}.
These peak values are within the range needed to perform experiments
associated with the results of Fig. \ref{fig_bi}. The effect can be enhanced
by stacking several layers of graphene together, separated from each other
 by a boron nitride spacer (rather than using multilayer graphene sheets).

\section*{Acknowledgements}
We are grateful to D. K. Ferry and A. A. Ignatov for sharing  their insights on the early developments of Boltzmann transport theory for semiconductors, and we thank N. A. Mortensen for useful remarks.
This work was partially
supported by the FEDER COMPETE Program and by the Portuguese Foundation for Science
and Technology (FCT) through grant PEst-C/FIS/UI0607/2013. We acknowledge
 support from the EC under Graphene Flagship (contract no. CNECT-ICT-604391).  The Center for
 Nanostructured Graphene (CNG) is sponsored by the Danish National Research 
Foundation, Project No. DNRF58. 
JES's work contract is financed in the framework of the Program of 
Recruitment of Post Doctoral Researchers for
the Portuguese Scientific and Technological System, 
with the Operational Program Human Potential (POPH) of the
QREN, participated by the European Social Fund (ESF) and national 
funds of the Portuguese Ministry
of Education and Science (MEC).

\begin{appendix}
\section{Exact solution of the Boltzmann equation}
\label{sec:AppA}

Here we give, for completeness,  a derivation of the exact solution for the
 relaxation-time Boltzmann equation
with uniform, time-dependent fields.  This situation has been analyzed by a 
large number of researchers in the past.  The solution is
implicit (but not explicitly stated) in the early work of  
Chambers\cite{Chambers}, and analyzed in detail by Ignatov and
 Romanov in their discussion of nonlinear electromagnetic 
properties of semiconductor superlattices\cite{Ignatov}.   
To solve (\ref{BE}) we proceed as follows.
Making the transformation
\begin{equation}
 f({\mathbf k},t)=e^{-t/\tau}g(\mathbf k,t)\,,
\end{equation}
Eq.(\ref{BE}) reads
\begin{equation}
 \tau \frac{\partial g({\mathbf k},t)}{\partial t} -k_0(t)
\frac{\partial g({\mathbf k},t)}{\partial k_x}=f_0e^{t/\tau}\,,
\end{equation}
where $k_0(t)=eE(t)\tau/\hbar$.
This differential equation can be solved by the method of characteristics. We thus  write
\begin{equation}
 \frac{dt}{\tau} =-\frac{dk_x}{k_0(t)}=\frac{dg({\mathbf k},t)}{f_0e^{t/\tau}}\,.
\end{equation}
The characteristic curves are defined by the solution of
\begin{equation}
 \frac{dt}{\tau} =-\frac{dk_x}{k_0(t)}\Leftrightarrow
k_0(t)dt=-\tau  dk_x\,,
\end{equation}
which upon integration gives
\begin{equation}
 \int^tk_0(t_1)dt_1+\tau k_x=C\,,
\end{equation}
which defines a family of curves for different $C$'s.
We can again use the characteristic relations and write
\begin{equation}
 dg({\mathbf k},t) = f_0(k_x,k_y)e^{t/\tau}\frac{dt}{\tau}\,.
\end{equation}
Using the equation for the characteristic curve we write
\begin{equation}
 dg({\mathbf k},t) = f_0[C/\tau-\int^tk_0(t_1)dt_1/\tau,k_y]e^{t/\tau}\frac{dt}{\tau}\,,
\end{equation}
which upon integration gives
\begin{equation}
 g({\mathbf k},t) = \int_{t_0}^t \frac{dt'}{\tau}
f_0[C/\tau-\int^{t'}k_0(t_1)dt_1/\tau,k_y]e^{t'/\tau}\,,
\end{equation}
and writing
\begin{equation}
C/\tau =  \int^tk_0(t_1)dt_1/\tau +k_x\,,
\end{equation}
the equation for $g({\mathbf k},t)$ reads
\begin{equation}
g({\mathbf k},t) = \int_{t_0}^t
\frac{dt'}{\tau}e^{t'/\tau}
f_0[k_x+\int^{t}_{t'}k_0(t_1)dt_1/\tau,k_y]\,,
\end{equation}
from which $f(\mathbf k,t)$ follows.
The value of $t_0$ is determined from the condition:
if $k_0(t)\rightarrow 0$ then $f(\mathbf k,t)\rightarrow f_0(\mathbf k)$.
In this limit we obtain
\begin{equation}
\lim_{E(t)\rightarrow 0}f(\mathbf k,t)
\rightarrow f_0(\mathbf k) e^{-t/\tau}\int_{t_0}^te^{t'/\tau}dt'/\tau\,,
\end{equation}
which implies that $t_0=-\infty$. Thus

\begin{equation}
f(\mathbf k,t) = e^{-t/\tau}\int_{-\infty}^t\frac{dt'}{\tau}e^{t'/\tau}
f_0[k_x+\int^{t}_{t'}k_0(t_1)dt_1/\tau,k_y]\,,
\label{BeqSol1}
\end{equation}
the result presented in the main text.

\section{Iterative solution of the Boltzmann equation}
\label{sec:AppB}

The results obtained in the bulk of the text for
the non-linear current can also be derived, although in a less elegant way, by an iterative approach.
We give here the derivation of the current $j_x^{3,\omega}$ for the case of graphene.
We assume a momentum independent relaxation time, but the method works as well if $\tau$
is momentum dependent.

Within the relaxation time approximation, Boltzmann equation reads
\begin{equation}
 \frac{\partial f}{\partial t}-\frac{e}{\hbar}\vec E\cdot\vec\nabla_{\vec k}f=-\frac{f-f_0}{\tau}
\end{equation}
where $e>0$, $f_0$ is the distribution function in equilibrium, and $f$ is the distribution function
in the presence of the field (that it, out of equilibrium). We assume that the system is subjected to
a finite AC field of the form
\begin{equation}
 \vec E=\epsilon_0\hat u_xe^{-i\omega t}+\epsilon_0^\ast\hat u_xe^{i\omega t}\,,
\end{equation}
where at some point in the calculation we take $\epsilon_0=\epsilon_0^\ast=E_0/2$.
We seek a distribution function in the form
\begin{equation}
 f(t)=f_0+f_1(t)+f_2(t)+f_3(t)\,,
\label{eq_f_expan}
\end{equation}
where the sub-index refers to the power of the field within the term of the distribution.

We note in passing that the solution of a differential equation of the form
\begin{equation}
 \dot y+ay=s(t)\,,
\end{equation}
where $s(t)$ is a source term, reads
\begin{equation}
 y(t)=e^{-a t}\int_{-\infty}^t s(t')e^{at'}dt'\,.
\label{eq_yt}
\end{equation}
For sure, this is indeed a particular solution, but one
 where the memory of the transient response has been lost;
this is assured by taking $t'=-\infty$ in the lower limit of the integral. In the context of the
response of an electron gas to an AC electric field, where dissipation exists, this choice for the lower
limit of the integral is physically justified.

We now plug in the expansion (\ref{eq_f_expan}) in Boltzmann equation and gather the terms with the
same order in the field. This leads to
\begin{eqnarray}
\label{eq_f1}
 \dot f_1 + \frac{f_1}{\tau}&=&\frac{e}{\hbar}\vec E\cdot\vec\nabla_{\vec k}f_0\,,\\
\label{eq_f2}
 \dot f_2 + \frac{f_2}{\tau}&=&\frac{e}{\hbar}\vec E\cdot\vec\nabla_{\vec k}f_1\,,\\
\label{eq_f3}
 \dot f_3 + \frac{f_3}{\tau}&=&\frac{e}{\hbar}\vec E\cdot\vec\nabla_{\vec k}f_2\,.
\end{eqnarray}
Equation (\ref{eq_f1}) is of the form (\ref{eq_yt}) and we obtain for $f_1$ the result
\begin{equation}
 f_1=\frac{\partial f_0}{\partial \epsilon}
\frac{e\vec v_F\cdot\hat u_x\epsilon_0}{1/\tau-i\omega}e^{-i\omega t}
+
\frac{\partial f_0}{\partial \epsilon}
\frac{e\vec v_F\cdot\hat u_x\epsilon_0^\ast}{1/\tau+i\omega}e^{i\omega t}\,,
\label{eq_f1_final}
\end{equation}
where $\epsilon=v_F\hbar k$ and $\vec v_F=v_F\vec k/k$. The details of the calculation are as
follows:
\begin{equation}
 f_1=e^{-t/\tau}\frac{e}{\hbar}\vec\nabla_{\vec k}f_0\cdot\int_{-\infty}^t
(\epsilon_0\hat u_xe^{-i\omega t}+\epsilon_0^\ast\hat u_xe^{i\omega t})
e^{t'/\tau}dt'\,.
\end{equation}
Upon integration, the result (\ref{eq_f1_final}) follows. We have also used the result
\begin{equation}
 \vec\nabla_{\vec k}f_0 = \frac{\partial f_0}{\partial \epsilon}\vec v_F\,.
\end{equation}
\begin{widetext}
We now proceed to the solution of equation (\ref{eq_f2}). Explicitly, we have
\begin{equation}
 \dot f_2 + \frac{f_2}{\tau} = \frac{e}{\hbar}
(\epsilon_0\hat u_xe^{-i\omega t}+\epsilon_0^\ast\hat u_xe^{i\omega t})
\cdot \vec\nabla_{\vec k}\left[
\frac{\partial f_0}{\partial \epsilon}
\left(
\frac{e\vec v_F\cdot\hat u_x\epsilon_0}{1/\tau-i\omega}e^{-i\omega t}
+\rm{H.\,c.}
\right)
\right]
\end{equation}
Taking $\epsilon_0=\epsilon_0^\ast$ and solving the differential equation, we obtain for
$f_2$ the result
\begin{eqnarray}
f_2&=& f_0''\frac{e^2v_F^2}{1/\tau-i\omega}(\vec v_F\cdot\hat u_x\epsilon_0)^2
\left(\frac{e^{-2i\omega t}}{1/\tau-2i\omega}+\tau\right)\nonumber\\
&+&f_0'\frac{e^2v_F}{\/\tau-i\omega}\frac{1}{\hbar k}
[\epsilon_0^2-(\epsilon_0\hat u_x\cdot\vec v_F/v_F)^2]
\left(\frac{e^{-2i\omega t}}{1/\tau-2i\omega}+\tau\right)+ \rm{H.\,c.}\,,
\end{eqnarray}
\end{widetext}
where
\begin{equation}
 f_0'=\frac{\partial f_0}{\partial \epsilon}
\end{equation}
and
\begin{equation}
 f_0''=\frac{\partial^2 f_0}{\partial \epsilon^2}\,,
\end{equation}
and the result ($\vec\epsilon_0=\epsilon_0\hat u_x$)

\begin{equation}
\vec\nabla_{\vec k}(\vec\epsilon_0\cdot\vec v_F)=  v_F\vec\nabla_{\vec k}(\vec\epsilon_0\cdot \vec k/k)
=v_F\left(
\frac{\vec \epsilon_0}{k}-\frac{\vec \epsilon_0\cdot\vec k }{k^3}\vec k
\right)
\end{equation}
has been used.
Clearly, $f_2$ does not contribute to the current, because
\begin{equation}
 \int_0^{2\pi}\cos\theta=\int_0^{2\pi}\cos^3\theta=0\,.
\end{equation}
We should note the presence in $f_2$ of a term that does not oscillate in time.
This term, however, will contribute to another term in $f_3$ oscillating with frequency
$\omega$.
Finally, we have to solve

\begin{equation}
 \dot f_3 +\frac{f_3}{\tau}=
\frac{e}{\hbar}
(e^{-i\omega t}+e^{i\omega t})\epsilon_0
\hat u_x\cdot \vec\nabla_{\vec k} f_2\,.
\label{eq_f3_dif_explicit}
\end{equation}
The rhs of the last equation together with its integration produces a number of terms.
We are interested in those terms proportional to $e^{-i\omega t}$.  We note that we can write
$f_2$ in form more convenient to our purposes (that is, power counting) as
\begin{eqnarray}
 f_2&=&\left[f_0''e^2(\vec v_F\cdot\hat u_x\epsilon_0)^2+
f_0'\frac{e^2 v_F}{\hbar k}[\epsilon_0^2-(\epsilon_0\hat u_x\cdot\vec v_F/v_F)^2]
\right]\nonumber\\
&\times&\left(
\frac{2}{(1/\tau)^2+\omega^2}+\frac{e^{-2i\omega t}}{(1/\tau-i\omega)(1/\tau-2i\omega)}
+\rm{H.\,c.}
\right)\,,\nonumber\\
\label{eq_f2_new_form}
\end{eqnarray}
where H.\,c. refers to the Hermitian conjugate of the second term.
Given the form of $f_3$ in (\ref{eq_f3_dif_explicit}) and equation (\ref{eq_f2_new_form})
it is a simple task to isolate those terms proportional
to $e^{-i\omega t}$; there are four such terms. The calculations are straightforward.
The result is
\begin{eqnarray}
f_3&=&\frac{e^3v_F^2}{\hbar}\epsilon_0\hat u_x\cdot\vec\nabla_{\vec k}
[f_0''(\epsilon_0\hat u_x\cdot\vec k/k)^2]g(\omega)\nonumber\\
&+&\frac{e^3v_F}{\hbar^2}\epsilon_0\hat u_x\cdot\vec\nabla_{\vec k}
[f_0'(\epsilon_0^2/k-(\epsilon_0\hat u_x\cdot\vec k)^2/k^3)]g(\omega)\nonumber\\
&+&\ldots\,,
\label{eq_f3_first_omega}
\end{eqnarray}
where $g(\omega)$ reads
\begin{eqnarray}
 g(\omega) &=& \frac{2}{1/\tau^2+\omega^2}\frac{e^{-i\omega t}}{1/\tau-i\omega}
\nonumber\\
&+&\frac{e^{-i\omega t}}{(1/\tau-i\omega)^2(1/\tau-2i\omega)}\,.
\end{eqnarray}
In Eq. (\ref{eq_f3_first_omega}) only the terms proportional to $e^{-i\omega t}$
are written explicitly.
The collisionless limit of $g(\omega)$ reads
\begin{equation}
 \lim_{\tau\rightarrow\infty}g(\omega)= \frac{3i}{2\omega^3}e^{-i\omega t}\,.
\end{equation}
We notice that the terms containing
derivatives of the $\delta-$functions do not contribute to the current. In this case,
the current that oscillates with frequency $\omega$ is simply given by
\begin{eqnarray}
 j^{(3,\omega)}_x&=&-\frac{e^4v_F}{\pi^2\hbar^3}\int_0^{2\pi}
\int_0^\infty kdk\cos^2\theta\delta(k-k_F)
\frac{1}{k^2}\times
\nonumber\\
&&
3\epsilon_0^3\sin^2\theta
g(\omega)\,.
\end{eqnarray}
Performing the integrations and writing $\epsilon_0=E_0/2$ we obtain
 \begin{equation}
   j^{(3,\omega)}_x=-\frac{3}{4}\frac{e^4}{\pi\hbar^3}\frac{v_F}{k_F}\frac{E_0^3}{8}g(\omega)\,,
 \end{equation}
which in the collisionless limit reads
\begin{equation}
 j^{(3,\omega)}_x=-i\frac{9}{8}\frac{e^4}{\pi\hbar^3}\frac{v_F}{k_F}\frac{E_0^3}{8}e^{-i\omega t}\,.
\end{equation}
The explicit form of $f_3$ is obtained from
\begin{eqnarray}
 f_3&=&f_0''' e^3(\vec\epsilon_0\cdot\vec v_F)^3g(\omega)+f_0''\frac{e^3}{\hbar}
\vec\epsilon_0\cdot\vec\nabla_{\vec k}(\vec v_F\cdot\vec\epsilon_0)^2g(\omega)\nonumber\\
&+&f_0''\frac{e^3}{\hbar}v_F\vec v_F\cdot\vec\epsilon_0
\left(\frac{\epsilon_0^2}{k}- \frac{(\vec\epsilon_0\cdot\vec k)^2}{k^3} \right)g(\omega)\nonumber\\
&+& f_0' \frac{e^3}{\hbar^2}v_F\vec\epsilon_0\cdot\vec\nabla_{\vec k}
\left(\frac{\epsilon_0^2}{k}- \frac{(\vec\epsilon_0\cdot\vec k)^2}{k^3} \right)g(\omega)\nonumber\\
&+& \ldots
\,,
\label{eq_f3_full}
\end{eqnarray}
where the following relations are useful
\begin{eqnarray}
\vec\epsilon_0\cdot\vec\nabla_{\vec k}(\vec\epsilon_0\cdot\vec v_F/v_F)^2 &=&
2(\vec\epsilon_0\cdot\vec v_F/v_F)
\times\nonumber\\
&&
\left(
\frac{\epsilon_0^2}{k}-\frac{(\vec\epsilon_0\cdot \vec k)^2}{k^3}
\right)
\end{eqnarray}
and
\begin{eqnarray}
\vec\epsilon_0\cdot\vec\nabla_{\vec k}
\left(
\frac{\epsilon_0^2}{k}-\frac{(\vec\epsilon_0\cdot \vec k)^2}{k^3}\right)
&=&-3\frac{\epsilon_0^2}{k^2}(\vec\epsilon_0\cdot\vec v_F/v_F)
\nonumber\\
&+&
3\frac{(\vec\epsilon_0\cdot\vec v_F/v_F)^3}{k^2}\,.
\end{eqnarray}
We also note the result
\begin{equation}
 \int F(k)\delta^{(n)}(k-k_F)dk=(-1)^nF^{(n)}(k_F)\,,
\end{equation}
where the superscript $(n)$ refers to the order of the derivative in order to $k$.
This result is used to prove  that the terms proportional to derivatives of
the $\delta-$function (the first three terms)
in Eq. (\ref{eq_f3_full}) give a zero contribution to the current.

\section{Details of calculation of the current in bilayer and trilayer graphene}
\label{sec:AppC}

Two parabolic bands characteristic of bilayer graphene lead to the following integrals entering the expression for the current density (\ref {eq:jx2}):
\begin{eqnarray}
\nonumber
& &I_{2n}(\kappa)=\int dk_xdk_yk_x\times\\
& &\theta\left\{\epsilon_F-\delta_{2,n}t_\perp-\frac{v_F^2\hbar^2\left[(k_x+\kappa)^2+k_y^2\right]}{t_\perp}\right\}\,,
\label{I2n}
\end{eqnarray}
In order to evaluate these integrals, we perform the same substitution as in the calculation of $I_{11}$, $k_x+\kappa=\tilde k_x$ and $k_y=\tilde k_y$. Thus, (\ref{I2n}) takes the form
\begin{eqnarray}
\nonumber
& &I_{2n}(\kappa)=\int d\tilde k_xd\tilde k_y (\tilde k_x-\kappa)\times\\
\nonumber
& &\theta\left(\epsilon_F-\delta_{2,n}t_\perp-\frac{v_F^2\hbar^2(\tilde k_x^2+\tilde k_y^2)}{t_\perp}\right)\\
& &=-\kappa\pi\frac{\left(\epsilon_F-\delta_{2,n}t_\perp\right)t_\perp}{v_F^2\hbar^2}\theta\left(\epsilon_F-\delta_{2,n}t_\perp\right)\,.
\end{eqnarray}

In the case of trilayer graphene, the current density is:
\begin{eqnarray}
& &j_x=-\frac{\sqrt{2}ev_F^2\hbar}{\pi^2t_\perp}e^{-t/\tau}\int^t_{-\infty}\frac{dt'}{\tau}e^{t'/\tau}\left[I_{31}(\kappa)+I_{33}(\kappa)\right]-\nonumber\\
& &-\frac{ev_F}{\pi^2}e^{-t/\tau}\int^t_{-\infty}\frac{dt'}{\tau}e^{t'/\tau}I_{32}(\kappa),\label{eq:jx3}
\end{eqnarray}
where
\begin{eqnarray}
& &I_{3n}(\kappa)=\int dk_xdk_y k_x\nonumber\times\\
& &\theta\left\{\epsilon_F-\delta_{3,n}\sqrt{2}t_\perp-\frac{v_F^2\hbar^2\left[(k_x+\kappa)^2+k_y^2\right]}{\sqrt{2}t_\perp}\right\}\,.
\end{eqnarray}
for $n=1,3$ and $I_{32}(\kappa)=I_{11}(\kappa)$ [see Eq. (\ref{eq:I11})]. Using this and the similarity between the integrals $I_{31}$, $I_{33}$ and $I_{21}$, $I_{22}$ (replacing $t_\perp  \to \sqrt{2}t_\perp$),
we obtain the final expression for the current density given in the text.

\end{appendix}


\end{document}